%% file: main.tex
\documentclass[conference]{IEEEtran}
\IEEEoverridecommandlockouts
\usepackage{cite}
\usepackage{amsmath,amssymb,amsfonts}
\usepackage{algorithmic}
\usepackage{graphicx}
\usepackage{textcomp}
\usepackage{xcolor}
\usepackage{booktabs}
\usepackage{times}
\usepackage{mathptmx}
\usepackage{helvet}
\usepackage{courier}
\usepackage{url}
\usepackage{subcaption}
\usepackage{hyphenat}
\usepackage{hyperref}
\usepackage{fancyhdr}
\usepackage{makecell}
\usepackage{multirow}
\usepackage{multicol}
\def\BibTeX{{\rm B\kern-.05em{\sc i\kern-.025em b}\kern-.08em
    T\kern-.1667em\lower.7ex\hbox{E}\kern-.125emX}}

\usepackage{fancyhdr}
\begin{document}

\title{
Policy-Guided ML for Energy Savings: Cell On/Off Switching under Operator QoS Constraints in Real 5G Networks
\\
\thanks{This work is supported by the Grant TRAINER-6G (PID2023-146748OB-I00) funded by MCIN/AEI/10.13039/501100011033 and by ERDF/EU, by the Smart Networks and Services Joint Undertaking (SNS JU) under the European Union’s Horizon Europe research and innovation programme under Grant Agreement No 101097083, BeGREEN project, and by the Open Call ECO-RAN under the EU-funded project 6G-SANDBOX (Grant Agreement No. 101096328).}
}

\author{%
\IEEEauthorblockN{David Reiss}
\IEEEauthorblockA{\textit{Signal Theory and Communications} \\
\textit{UPC, ETSETB}\\
Barcelona, Spain\\
david.reiss@upc.edu}
\par\vspace{1.5ex}
\IEEEauthorblockN{Daniel Camps-Mur}
\IEEEauthorblockA{\textit{Mobile Wireless Internet Group} \\
\textit{i2cat}\\
Barcelona, Spain \\
daniel.camps@i2cat.net}
\and
\IEEEauthorblockN{Miguel Catalan-Cid}
\IEEEauthorblockA{\textit{Mobile Wireless Internet Group} \\
\textit{i2cat}\\
Barcelona, Spain \\
miguel.catalan@i2cat.net}
\par\vspace{1.5ex}
\IEEEauthorblockN{Oriol Sallent}
\IEEEauthorblockA{\textit{Signal Theory and Communications} \\
\textit{UPC, ETSETB}\\
Barcelona, Spain \\
jose.oriol.sallent@upc.edu}
}

\maketitle
\thispagestyle{fancy} 

\begin{abstract}
Energy efficiency is a critical concern in the deployment and operation of 5G networks, particularly due to the low utilization of 4G and 5G carriers during off-peak hours. While considerable research has focused on designing energy-efficient cell on/off switching strategies that avoid disrupting user connectivity, the integration of operator-specific policies to guarantee particular Quality of Service (QoS) levels has received limited attention. This paper presents a machine learning (ML)-based energy saving strategy, trained using a real-world dataset from a European mobile operator, that enforces operator-defined policies that jointly consider strong throughput requirements and maximum outage tolerance constraints. By tuning the model’s class ratios during training, the proposed solution enables operators to manage the trade-off between energy savings and QoS policy compliance  prior to deployment in live networks. Evaluation results show that the method provides substantial energy savings while maintaining policy-compliant service levels under realistic 5G operating conditions. 
\end{abstract}

\input{introduction}

\input{dataset}
\input{classifier_design}

\input{classifier_eval}

\input{conclusions}

\bibliographystyle{IEEEtran}
\bibliography{references}

\end{document}

%% file: introduction.tex
\section{Introduction}\label{sec:intro}

\begin{figure*}[t]
    \centering

    \begin{subfigure}[b]{\textwidth}
        \centering
        \includegraphics[width=0.98\textwidth]{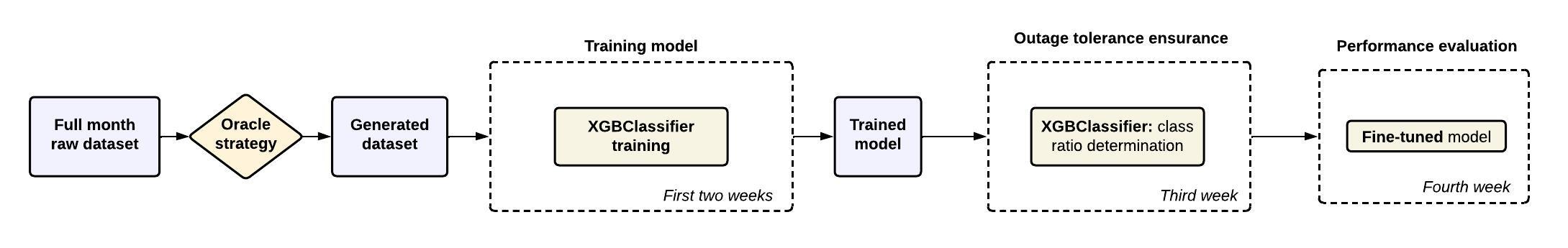}
    \end{subfigure}

    \caption{High-level schematic description.}
    \label{fig:high-level-scheme}
\end{figure*}

{With the global deployment of 5G networks, energy consumption has emerged as a critical concern for Mobile Network Operators (MNOs). The increased demand for high-capacity and low-latency applications has driven a more extensive usage of the spectrum resources, leading to a denser Radio Access Network (RAN). In 5G and beyond networks, this radio densification is characterized by the integration of higher frequency carriers, which complement traditional macro cells to boost capacity. These high-capacity cells employ techniques such as Massive Multiple Input Multiple Output (MIMO), which lead to a significant increase in energy consumption. Addressing this issue is essential to ensure the sustainable operation of future mobile networks. }

{Energy saving techniques in the RAN often rely on the selective deactivation of specific network elements. However, to ensure practical applicability, it is essential to consider the impact of such strategies on the Quality of Service (QoS). Incorporating these performance indicators is critical for enhancing energy efficiency without compromising the end-user experience. Existing energy savings studies include heuristic or rule-based methods \cite{sleep-modes}, AI-driven strategies \cite{mswim-paper}, QoS-aware techniques \cite{advanced-sleep-modes}, and even some approaches that use open datasets\cite{gnn-switching}. However, most evaluate energy savings against minimal or sufficient QoS requirements and do not offer practical tools for MNOs to configure energy-saving strategies under explicit QoS constraints. }

{The main contribution we present is the development of an ML-driven cell on/off switching strategy that enhances control over the trade-off between energy savings and provided QoS. We explore different dimensions of QoS constraints to fully characterize the performance of our solution, showing how different requirements influence the achievable savings. This work builds on top of the study we performed in \cite{previous-paper}, where we utilized a dataset from an European MNO to analyze the 5G cells' switch off opportunities, as well as to quantify the energy saving QoS trade-off. By running an oracle strategy (not practicably implementable) we searched for periods of time on which the 5G traffic could be offloaded to the 4G cells found in the same site and sector. Moreover, a regression based method allowed us to estimate the throughput provided per UE in the 4G cells during the switch off periods. The dataset, which covers an extensive area of an European city and its surroundings, has a granularity of 15 minutes, and spans one full month, is also used in this paper to train and evaluate the performance of the proposed ML-driven strategy. However, to limit the computational cost of training and optimization, we target a subset of 70 cells from the 200 cells we analyzed in the previous work. Those are located in an urban scenario and have a high diversity of switch off opportunities, therefore presenting a strong interest for ML-driven strategies to be tested. }

{The paper is structured as follows. The next section presents the QoS policies definition. Section \ref{sec:solution-design}, presents the design of ML-driven strategies to enforce the QoS constraints. Then, Section \ref{sec:evaluation} presents the performance evaluation. Finally, in Section \ref{sec:conclusions}, we present the conclusions and the future work.}

%% file: dataset.tex
\section{QoS-guided ES policies definition}\label{sec:use-case}

{The solution proposed in this paper is designed to enable network operators to improve energy efficiency while adhering to defined QoS requirements. Those can span a variety of performance metrics, including handover success rates, data throughput, dropped user sessions, and user-experienced delays. Specifically, this work focuses on two key dimensions of QoS enforcement, which jointly define the requested operator QoS policy: target throughput level and outage tolerance constraints.}

{First, specifying a target throughput level to be ensured in the 4G cells during the switch-off periods is crucial for the energy saving mechanism to adapt to different network conditions and service demands. In this work we only focus in one throughput level which is considered to be strong according to popular video streaming applications: 15 Mbps. Concretely, this would allow the end user to visualize HD 1080p YouTube\footnote{\url{https://support.google.com/youtube/answer/78358?hl=en}} videos, and UHD 4K Netflix\footnote{\url{https://help.netflix.com/en/node/306}} movies.  

Second, since the proposed ML-driven solution is intended for deployment in real-world across dynamic environments with unseen data, it is critical to consider outage decisions, i.e., incorrect switch-off choices that lead to service degradation. To systematically assess and mitigate this risk, we also introduce an outage tolerance constraint to be imposed: C, which is defined as: }

\begin{flushleft}
    \begin{equation}
    C : \frac{N_{out}}{N_{on}} = \frac{\sum_{t=0}^{T} f_1(o(t), s(t))}{\sum_{t=0}^{T} f_{1'}(o(t))} \leq \gamma
    \end{equation}
\end{flushleft}

where, 

\begin{flushleft}
    \begin{equation}
    f_1(x, y) = 
    \begin{cases}
    1, & \text{if } (x, y) = (1,0) \\
    0, & \text{otherwise}
    \end{cases}
    \end{equation}
\end{flushleft}

\begin{flushleft}
    \begin{equation}
    f_{1'}(x) = 
    \begin{cases}
    1, & \text{if } x = 1 \\
    0, & \text{otherwise}
    \end{cases}
    \end{equation}
\end{flushleft}

$N_{out}$ is the number of outage decisions of the evaluated strategy, and $N_{on}$ represents the total ON decisions taken by the oracle strategy. $T$ is the discretized time period under analysis (due to the fifteen minute granularity of the dataset), and $o(t)$ and $s(t)$, are the decisions taken by the oracle strategy and the proposed strategy (defined in Section \ref{sec:solution-design}) at each time interval $t$, respectively. C bounds the outage decisions normalized to the total ON decisions to a defined threshold $\gamma$.



%% file: classifier_design.tex
\section{Design of ML-driven policy enforcement strategies}\label{sec:solution-design}

{This section introduces the design of the proposed ML-driven strategy. Subsection A demonstrates how the proposed model is aware of the target throughput level, and subsection B presents the mechanism to enforce the outage tolerance constraints.}

\subsection{Training dataset generation for target throughput awareness}

{Since the strategy needs to decide the 5G cells' on/off switching decision, we formulate the problem as a binary classification task. The underlying model is the XGBoost Classifier, which is based on a tree boosting algorithm \cite{xgboost-reference}, and has been widely proven to effectively solve classification and regression tasks. As depicted in Figure \ref{fig:high-level-scheme}, to train the models we generated a dataset by running the oracle strategy for the 15 Mbps throughput level over two weeks of data. This allowed us to collect information on the switch-off periods and the throughput estimation across the 70 analyzed cells. The training dataset format is illustrated in Table \ref{tab:dataset}.} 

\begin{table}[h]
\centering
\caption{Training Dataset}
\begin{tabular}{|c|c|c|c|c|c|}
\hline
\makecell{Time} & \makecell{5G\\ cell\\ load \\(\%)} & \makecell{4G \\cell 1 \\load \\(\%)} & {···} & \makecell{4G \\cell 5 \\load \\(\%)} & \makecell{Decision \\ (15 Mbps)}\\
\hline
$t_1$ & 54.3 & 45.2 & {···} & 67.8 & 1 \\
\hline
$t_2$ & 37.8 & 40.1 & {···} & 50.1 & 1 \\
\hline
$t_3$ & 24.2 & 35.1 & {···} & 42.5 & 0 \\
\hline
\end{tabular}
\label{tab:dataset}
\end{table}

{It is important to mention that the dataset could be generated for any throughput level (e.g., 5 Mbps, 10 Mbps, 20 Mbps, 25 Mbps, etc.), with the model subsequently re-trained to reflect this configuration. This indicates that the model’s awareness of the target throughput level is inherently determined by the dataset generation process and, therefore, can be adapted to ensure different throughput requirements. }

{The inputs to the model are the target 5G cell load, and the load of the five active 4G cells of the same site and sector. The labeled outputs are the last six columns, which represent with a "0" if the 5G cell can be switched off while maintaining the corresponding throughput level, and with a "1" if the 5G cell must be on. Note that, although having access to a large collection of KPIs (in the MNO's dataset), we used a limited set of input features to reduce the complexity of the model, obtaining a significant accuracy while reducing the model computational and energetic footprint. To mimic the oracle strategy, the model is trained to decide the on/off switching for the next fifteen-minute interval based on the current load conditions, i.e., in a predictive way. In addition, the cell name is not provided, allowing our model to be applied to unseen ones.}

\subsection{Class ratio tuning for outage tolerance enforcement}

{In binary classification tasks, two type of errors are encountered: false positives (FP), and false negatives (FN). Within the context of our use case, FPs correspond to missed switch-off opportunities, which can lead to energy saving wasted decisions. Conversely, FNs are identified as outage decisions, which can lead to degraded user experience. The class ratio hyperparameter (defined as \textit{scale pos weight} in XGBoost documentation\footnote{\url{https://xgboost.readthedocs.io/en/stable/index.html}}) allows for the model to weight the output classes (in our case "0" or OFF, and "1" or ON). It is typically used to control the balance of positive and negative weights, and adjusting it becomes useful to deal with unbalanced class distributions, i.e., large difference between the number of positive and negative instances. In our case, tuning the class ratio hyperparameter allows us to enhance control over both type of erroneous on/off decisions, which becomes key for our strategy to enforce the second QoS dimension: outage tolerance constraints (equivalent to biasing the model towards positive instances, i.e., switching on the 5G cell).  }

\begin{figure}[h]
    \centering
    \includegraphics[width=0.49\textwidth]{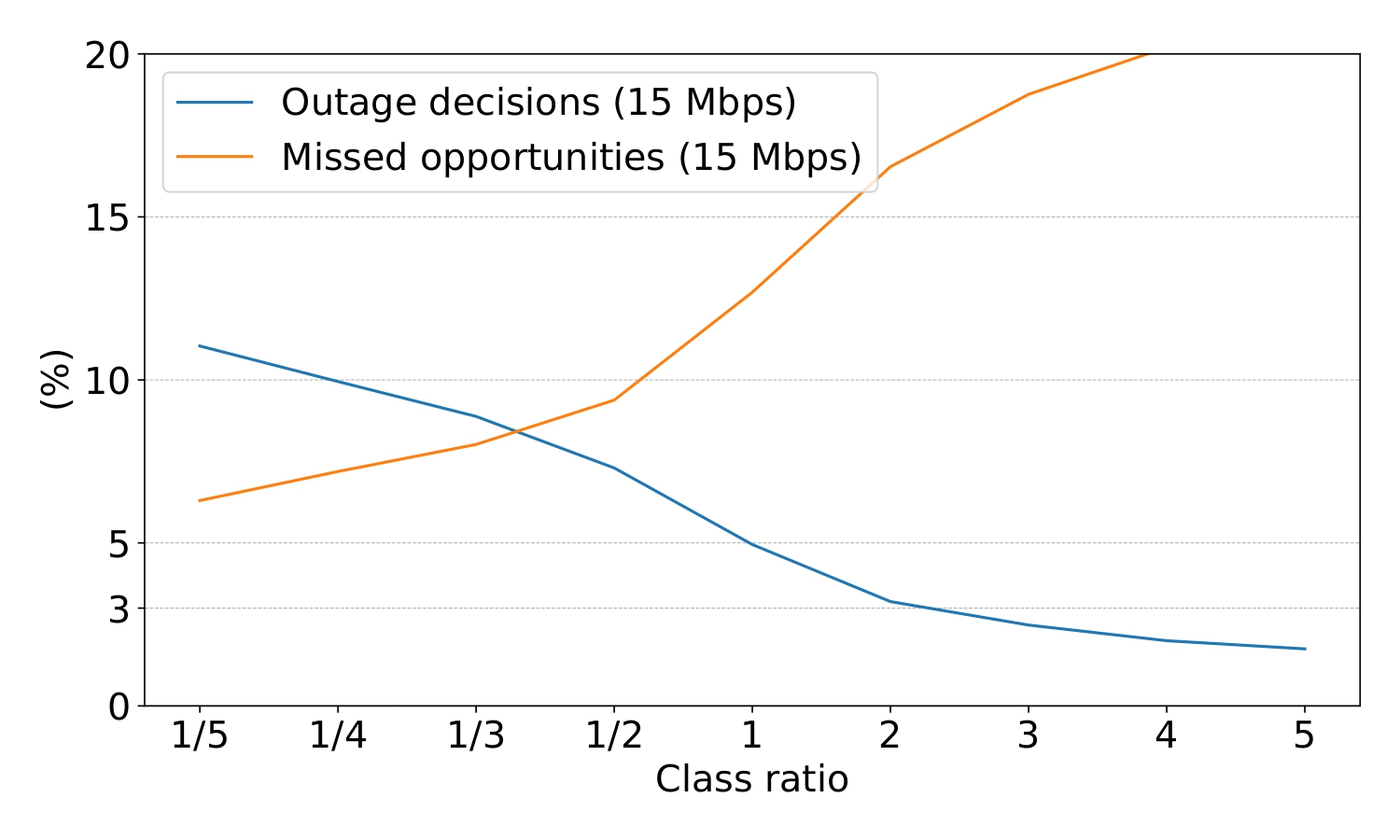}
    \caption{Class ratio effect over outage decisions (left) and missed opportunities (right).}
    \label{fig:missed-outage}
\end{figure}

{To test the effect of tuning the class ratio hyperparameter, we train several models with different class ratio values. Then we use the third week of the generated dataset to test the effect of those (Figure \ref{fig:high-level-scheme}). Figure \ref{fig:missed-outage}, illustrates the effect of different class ratio settings into the performance of the model in terms of outage decisions and missed opportunities normalized to the total ON and OFF decisions, respectively. }

{We observe that outage decisions decrease as the class ratio is increased, while missed opportunities increase. This means that imposing outage tolerance constraints will inevitably make the model to loose switch-off opportunities, therefore reducing the achievable savings. Moreover, the outage decisions and missed opportunities cross-point also provides significant information. It represents the class ratio which balances the model to the data distribution, i.e., the one that minimizes outage decisions and missed opportunities at the same time. As we quantified in \cite{previous-paper}, the energy savings-QoS trade-off demonstrates that as higher the throughput requirement, lower the switch-off opportunities. This translates into the generated dataset having less "0"s (OFFs) than "1"s (ONs), and that is why the cross-point is found to be shifted towards the left side of the plot. }

%% file: classifier_eval.tex
\section{Performance evaluation}\label{sec:evaluation}

\begin{figure*}[t]
    \centering

    \begin{subfigure}[b]{\textwidth}
        \centering
        \includegraphics[width=0.32\textwidth]{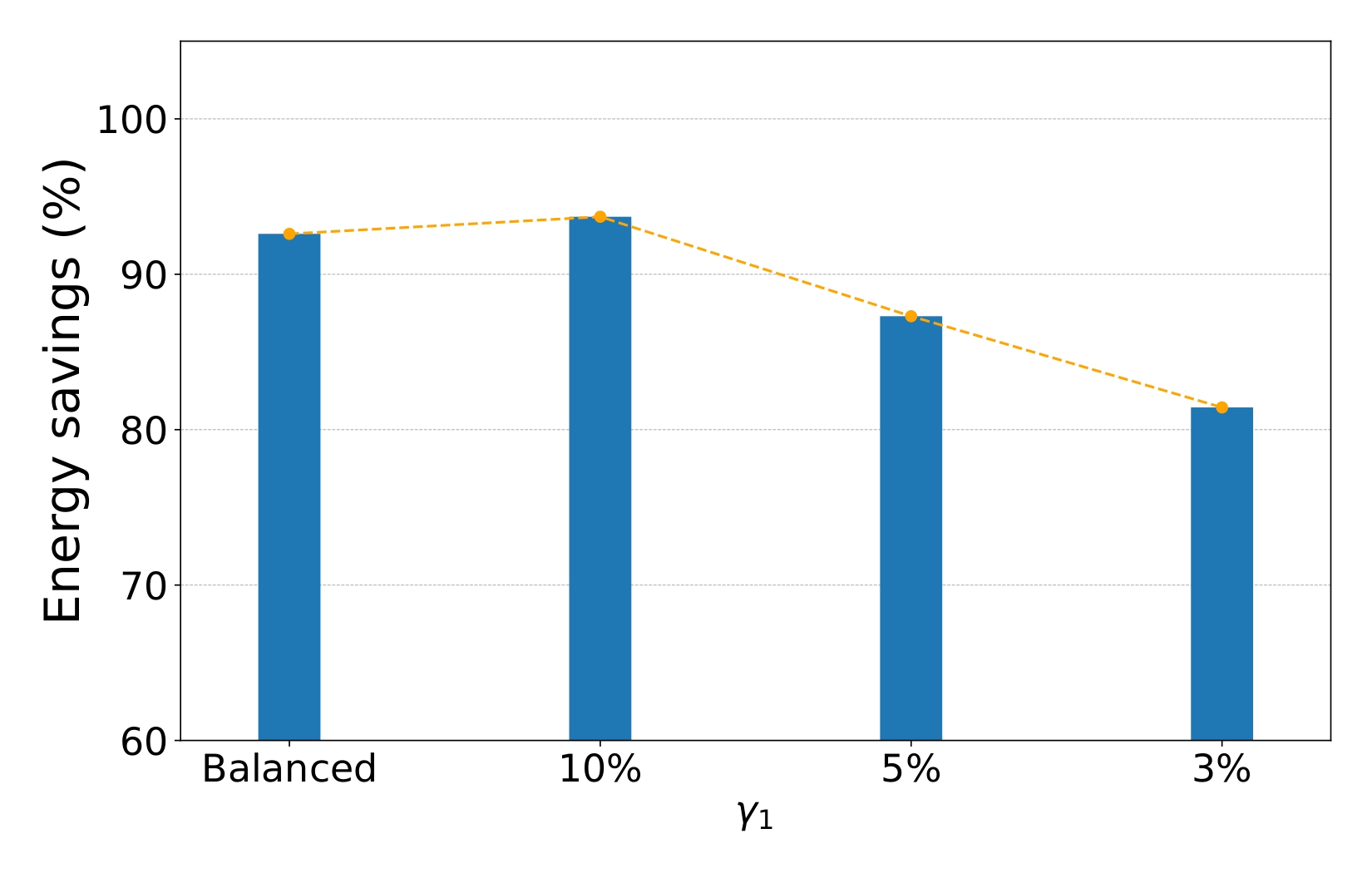}
        \hfill
        \includegraphics[width=0.32\textwidth]{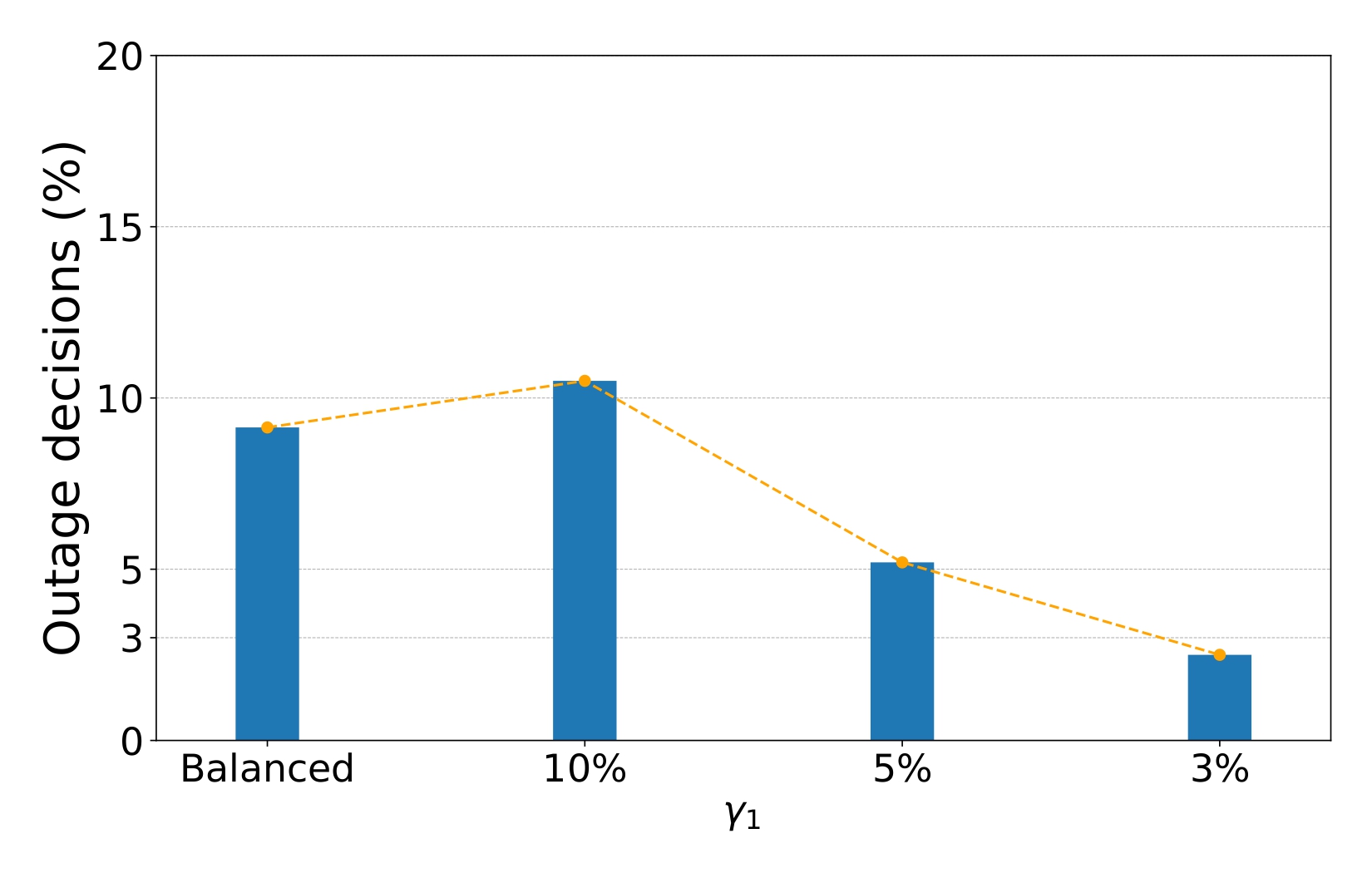}
        \hfill
        \includegraphics[width=0.32\textwidth]{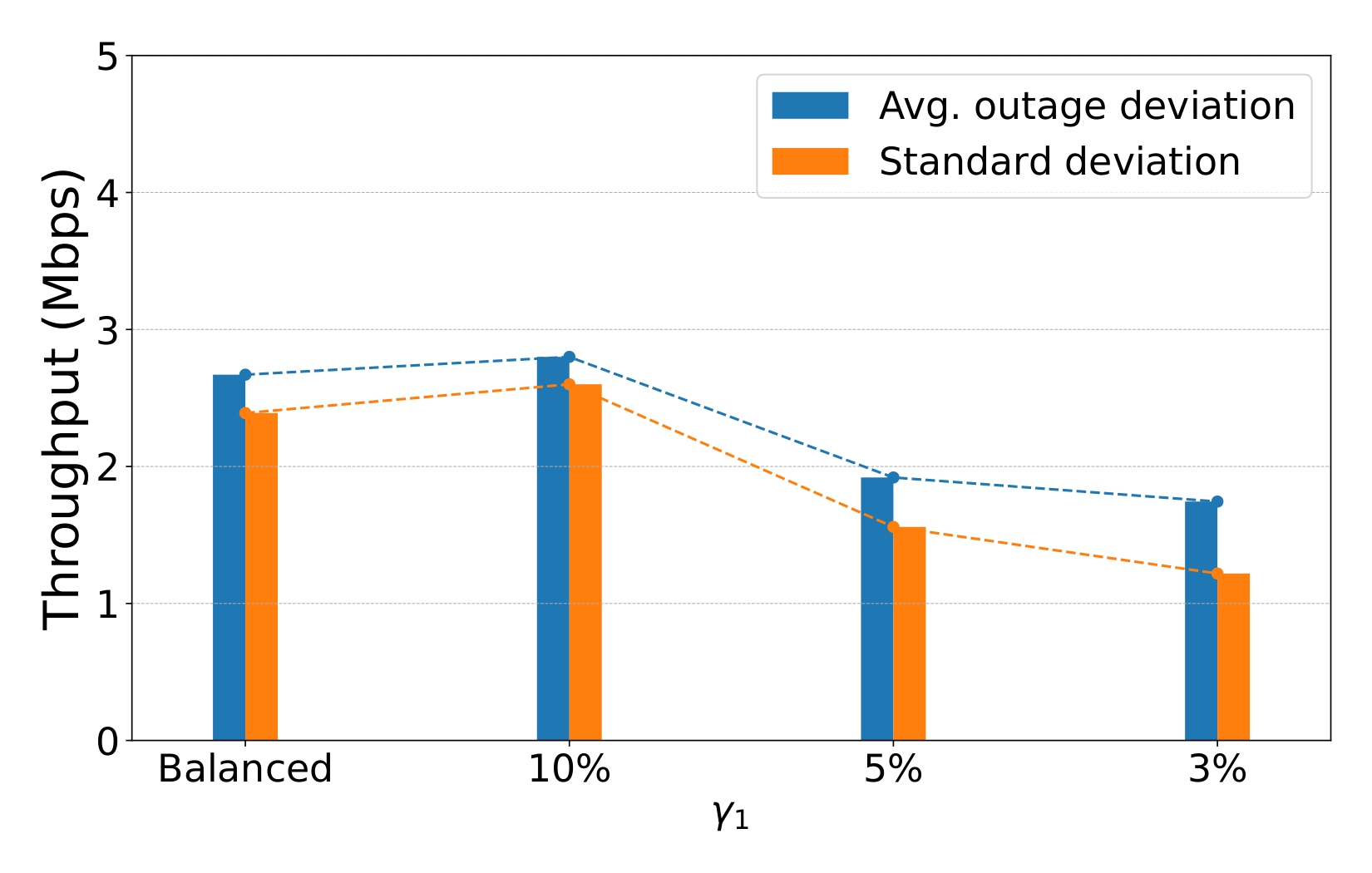}
    \end{subfigure}
    \caption{Energy savings (left), outage decisions (middle), and outage deviation (right) over evaluation week.}
    \label{fig:overall}
\end{figure*}

{To provide robust and reliable results, we evaluate the performance of the classifier over the fourth week of the generated dataset, which is unseen for the model (Figure \ref{fig:high-level-scheme}). Moreover, we demonstrate that tuning the class ratio is a valid tool to bound the outage decisions the model takes. To do so, we will test three different $\gamma$ values: 10\%, 5\%, 3\%. The methodology is straightforward: we select the class ratio which, according to the testing week results, ensures outage decisions will be below the $\gamma$ constraint. Thus, the class ratios to be used are 1/4, 1, and 3, respectively. Recall that, to maximize energy savings we must select the lowest class ratio which ensures the outage constraint. In addition, we also provide the results obtained when a balanced model is used, i.e., we minimize outage decisions and missed opportunities at the same time. This model is trained with a class ratio equal to 1/3 (cross-point of Figure \ref{fig:missed-outage}).}

{The evaluation metrics consist on: (i) the achieved energy savings, when compared to the optimal solution (oracle strategy), (ii) the outage decisions, to prove compliance with the outage constraint, and (iii) the outage deviation, which is defined as the error between the throughput estimation (15 Mbps) and the real throughput provided when an outage occurs. The latter allows us to quantify the severity of the outage decisions.}

{Figure \ref{fig:overall} shows the results obtained from the proposed model over the evaluation week. The balanced model has obtained a 92.6\% of savings with respect to the optimal solution. Moreover, the outage decisions have been equal to 9.1\% with an average outage error of 2.7 Mbps. Recall that, this results represent the case in which outage decisions and missed opportunities are minimized at once, i.e., any other class ratio will negatively impact either the outage decisions or the achievable savings.  }

{Analyzing the performance of the model under outage tolerance constraints, we first observe in the left figure that potential energy savings are achieved across all outage constraints. Even under the most stringent conditions (15 Mbps, $\gamma \leq 3\%$), the model performs approximately at 82\% from the optimal solution. However, the consequence is clear, as lower the outage tolerance, lower the energy savings. Second, we observe in the middle figure that outage decisions have been close to the imposed constraints, but violated the threshold at $\gamma = 10\%$, and $\gamma = 5\%$. Concretely, those were found to be equal to 10.5\% and  5.1\%, respectively. Although not an excessive error, this highlights the importance of implementing re-training schemes to adjust the model to varying network behavior from week to week (or other defined time period). Third, we observe in the right figure that employing different class ratios to reduce outage decisions has had a positive effect on outage deviation. Specifically, both the average deviation, and the standard deviation across cells, have been clearly reduced. This result implies that under high $\gamma$ requirements, the impact of outage decisions on the end user becomes less severe. For instance, at $\gamma = 3\%$, the average deviation was measured to be 1.75 Mbps with a standard deviation of 1.2 Mbps. Consequently, in outage scenarios, UEs rarely experienced data rates below 10 Mbps (15 Mbps - 1.75 Mbps – 2 x 1.2 Mbps).  }

{Finally, to provide a general perspective, Table \ref{tab:times} shows the average switch-off time across the 70 analyzed cells over the evaluation week. Additionally, to show the achievable energy savings under soft throughput requirements, we provide the results obtained when repeating the full workflow (Figure \ref{fig:high-level-scheme}) with a 5 Mbps requirement instead of 15 Mbps one. This demonstrates how both dimensions of QoS constraints impact the achievable savings.}

\begin{table}[h]
\centering
\caption{Average switch-off times across analyzed cells (\%)}
\begin{tabular}{|c|c|c|c|}
\cline{2-4}
\multicolumn{1}{c|}{} & {$\gamma = 10\%$} & {$\gamma = 5\%$} & {$\gamma = 3\%$} \\
\hline
5 Mbps & 73.1 & 67.9 & 57.2 \\
\hline
15 Mbps & 47.4 & 44.6 & 41.8 \\
\hline
\end{tabular}
\label{tab:times}
\end{table}

%% file: conclusions.tex
\section{Conclusions}\label{sec:conclusions}

{In this work, we have presented a novel ML-driven QoS-aware energy saving strategy based on 5G cell on/off switching and traffic offloading. Building on top of a realistic ES-QoS trade-off characterization based on real data, we have provided a robust solution to enhance control over energy savings and QoS policies across two different dimensions: target throughput requirements, and outage tolerance constraints. Moreover, in the performance evaluation, we demonstrated the achievable energy savings under soft and strong service requirements, which evidence that huge energy consumption reduction can be achieved in current cellular deployments.}

{Future research should aim at generalizing the proposed solution to any target throughput level, and extend outage tolerance constraints scope (e.g., introduce outage deviation constraints). Moreover, it must incorporate lightweight mechanisms for dynamically adjusting class ratios under evolving network conditions. A promising approach consists on integrating the proposed solution as an rApp in the Non-Real-Time RAN Intelligent Controller of the O-RAN architecture. By leveraging feedback from the RAN nodes, the system could implement an online learning framework in which class ratios are periodically adjusted according to the models performance and QoS policies.} 